\documentclass[11pt]{article}

\usepackage{authblk}
\usepackage{natbib}
\usepackage[a4paper, top=1.2in, bottom=1.5in, left=1in, right=1in]{geometry}
\usepackage{color}
\usepackage{amsmath}
\usepackage{amssymb}
\usepackage{amsthm}
\usepackage{bm}
\usepackage{subcaption}
\usepackage{ragged2e}
\usepackage{booktabs}
\usepackage{siunitx}  
\usepackage{float}    

\usepackage{pgf,tikz}
\usetikzlibrary{shapes.geometric, arrows, positioning, fit}
\tikzstyle{process} = [rectangle, minimum width=3cm, minimum height=1cm, text centered, draw=black]
\tikzstyle{arrow} = [thick,->,>=stealth]

\usetikzlibrary{arrows,shapes.arrows,shapes.geometric,shapes.multipart, decorations.pathmorphing,positioning,shapes.swigs,}

\tikzstyle{startstop} = [rectangle, rounded corners, minimum width=3cm, minimum height=1cm,text centered, draw=black]
\tikzstyle{process} = [rectangle, minimum width=3cm, minimum height=1cm, text centered, draw=black]
\tikzstyle{decision} = [diamond, minimum width=3cm, minimum height=1cm, text centered, draw=black, aspect=2]
\tikzstyle{arrow} = [thick,->,>=stealth]

\usepackage{algorithm}
\usepackage{algpseudocode}

\usepackage{array, caption, threeparttable}
\usepackage{url}
\usepackage{hyperref}
\usepackage{xcolor}
\hypersetup{
    colorlinks,
    linkcolor={blue},
    citecolor={blue},
    urlcolor={blue}
}

\usepackage{setspace}
\usepackage{appendix}

\newcommand{\PP}{\mathbb{P}}
\newtheorem{theorem}{Theorem}

\newtheorem{proposition}{Proposition}

\theoremstyle{definition}
\newtheorem{definition}{Definition}

\usepackage{xr}
\makeatletter
\newcommand*{\addFileDependency}[1]{
  \typeout{(#1)}
  \@addtofilelist{#1}
  \IfFileExists{#1}{}{\typeout{No file #1.}}
}
\makeatother

\newcommand*{\myexternaldocument}[1]{
    \externaldocument[si:]{#1}
    \addFileDependency{#1.tex}
    \addFileDependency{#1.aux}
}
\myexternaldocument{SI}

\makeatletter
\renewcommand*{\thesection}{\arabic{section}}

\renewcommand*{\p@subsection}{}

\renewcommand*{\p@subsubsection}{}
\makeatother

\usepackage[nameinlink,capitalise]{cleveref}
\crefname{section}{Sec.}{Secs.}
\Crefname{section}{Section}{Sections}
\crefname{figure}{Fig.}{Figs.}
\Crefname{figure}{Figure}{Figures}
\crefname{equation}{Eq.}{Eqs.}
\Crefname{equation}{Equation}{Equations}

\title{Peer-induced Fairness: A Causal Approach for Algorithmic Fairness Auditing}

\author[1]{Shiqi Fang\thanks{S.Fang-6@sms.ed.ac.uk}}
\author[1]{Zexun Chen\thanks{Zexun.Chen@ed.ac.uk}}
\author[1]{Jake Ansell\thanks{J.Ansell@ed.ac.uk}}

\affil[1]{\textit{Business School, University of Edinburgh, Edinburgh, EH8 9JS, United Kingdom.}}

\date{}

\begin{document}

\maketitle
\noindent
\begin{abstract}
With the European Union's Artificial Intelligence Act taking effect on 1 August 2024, high-risk AI applications must adhere to stringent transparency and fairness standards. This paper addresses a crucial question: how can we scientifically audit algorithmic fairness? Current methods typically remain at the basic detection stage of auditing, without accounting for more complex scenarios. We propose a novel framework, ``peer-induced fairness'', which combines the strengths of counterfactual fairness and peer comparison strategy, creating a reliable and robust tool for auditing algorithmic fairness. Our framework is universal, adaptable to various domains, and capable of handling different levels of data quality, including skewed distributions. Moreover, it can distinguish whether adverse decisions result from algorithmic discrimination or inherent limitations of the subjects, thereby enhancing transparency. This framework can serve as both a self-assessment tool for AI developers and an external assessment tool for auditors to ensure compliance with the EU AI Act. We demonstrate its utility in small and medium-sized enterprises' access to finance, uncovering significant unfairness—41.51\% of micro-firms face discrimination compared to non-micro firms. These findings highlight the framework's potential for broader applications in ensuring equitable AI-driven decision-making.

\end{abstract}


\section{Introduction}

\label{sec:introduction}
The increasing adoption of data-driven methodologies across various sectors has heightened global concerns about algorithmic bias. These concerns are underscored by the recent enactment of the European Union's Artificial Intelligence Act (EU AI Act), effective from 1 August 2024 \citep{madiega2021artificial}. The Act represents the world's first comprehensive legal framework for Artificial Intelligence (AI). It imposes stringent requirements on high-risk AI applications, such as credit scoring systems, mandating the rigorous identification and mitigation of discrimination risks. Additionally, the Act requires that these AI systems undergo thorough assessments both prior to deployment and continuously throughout their operational life cycle, thereby ensuring sustained compliance with transparency and fairness standards.

The implementation of the EU AI Act necessitates the urgent development of a universal and transparent tool capable of ensuring compliance with rigorous standards. This task is critical not only for regulators but for everyone involved in the AI community. A key challenge lies in scientifically assessing the fairness of decisions made by AI models—a question that remains at the forefront of ethical AI development.

Over the past decade, the importance of algorithmic fairness has increasingly gained recognition as a vital component of responsible technology use and ethical AI development. Despite this growing awareness, a substantial body of literature highlights the trade-offs between accuracy and fairness \citep{kim_within-group_2023, huang_fairness_2020, guldogan_equal_2022, dixon_measuring_2018, foulds_intersectional_2020, chen_measuring_2022, hickey_fairness_2020, dwork_fairness_2012, hardt_equality_2016}. However, there remains a significant gap in the development of robust auditing frameworks specifically designed to assess and monitor algorithmic bias. 
Some studies have proposed approaches for auditing algorithmic fairness with data-driven models \citep{cherian2024statistical, brundage2020toward, xue2020auditing, tramer2017fairtest, si2021testing, yan2022active}. However, these approaches primarily focus on fairness detection based on internally accessible data and models. When regulators obtain data externally, the underlying algorithmic models and decision-making processes are often unknown, necessitating a reassessment of whether the decisions produced by AI developers' models are fair. Besides, although identifying the presence of bias is necessary, it represents only the initial step in the comprehensive auditing of AI systems, and it is insufficient for ensuring robust and reliable auditing frameworks. The two key factors concerning the complexities inherent in the real world—\textbf{universality} and \textbf{transparency}—are often overlooked.

\textbf{With respect to universality}, a mature auditing framework should be adaptable to datasets with different characteristics. For example, a critical challenge in the development of auditing tools is the issue of data quality, particularly data scarcity and imbalance, which are pervasive in real-world datasets \citep{lessmann_benchmarking_2015, chen_interpretable_2024, sha_lessons_2023, dablain_towards_2022}. Historical biases frequently result in the under-representation of protected groups within datasets \citep{iosifidis_dealing_2018}. This imbalance, particularly the under-representation of minority groups in training data (i.e., representational disparity), leads to their diminished influence on model objectives and reduces their influence on model objectives \citep{hashimoto_fairness_2018}. Consequently, biased discrimination measures and unreliable audit outcomes may emerge \citep{sha_lessons_2023, dablain_towards_2022, sha_leveraging_2022, yan2020fair}. A particularly concerning issue arises when an individual or organisation is flagged as discriminated against in one dataset but deemed privileged in another due to varying degrees of imbalance. Such inconsistencies pose significant challenges to conducting universal audits. Current frameworks frequently assume that data is balanced across different groups, an assumption rarely met in practice. Some studies propose oversampling techniques, such as SMOTE, to address data imbalance \citep{sha_lessons_2023, sha_leveraging_2022, yan2020fair}. However, these methods can inadvertently ``smooth out''  critical features within the data, thereby altering the intricate relationships between variables \citep{chen_interpretable_2024}, which may ultimately lead to distorted auditing outcomes. Such flawed outcomes can be disastrous for regulators, companies, and third-party auditors, as they could introduce even greater risks. Audits based on unrealistic assumptions may fail to detect genuine instances of algorithmic discrimination, allowing unfair practices to persist. Furthermore, inaccurate audits could lead to misguided adjustments in algorithms, potentially worsening performance or introducing new biases, thus compounding the original issues. If stakeholders, including the public, perceive the auditing process as flawed or unreliable, it can erode confidence in both the regulators and the entities being audited. \textbf{In terms of transparency}, both regulatory authorities and researchers have consistently underscored the importance of transparent and explainable models that provide clear justifications for decision-making \citep{chen_interpretable_2024, voigt_eu_2017}. Current fairness-related criteria often incorporate explainability into the fairness assessment \citep{zhao2023fairness, hickey_fairness_2020, chen_measuring_2022}. However, a gap remains in delivering clear and understandable explanations. This gap is critical, as understanding the reasons behind rejections is essential for ensuring that decisions are perceived as fair. An opaque auditing framework can similarly lead both the audited entities and the public to perceive the audit process as incredible, thereby undermining its credibility and public confidence. Therefore, an effective auditing framework must elucidate the reasons for adverse decisions, particularly to distinguish whether such decisions are due to discrimination or inherent limitations.

We propose a reliable and robust audit framework that solves the above issues, embodying both universality and transparency. It is a causally-oriented approach to fairness \citep{kusner_counterfactual_2017}. Compared to traditional static fairness criteria, a causally-oriented approach offers a more effective means of addressing real-world challenges. This approach is particularly valuable for practitioners, policymakers, and judicial authorities tasked with implementing algorithms designed to mitigate discrimination. Selecting an appropriate fairness definition that aligns with the specific nuances of each scenario can be a significant challenge. For instance, the parameters for fairness in addressing gender disparities may differ substantially from those relevant to racial issues, or from considerations in broader, non-demographic contexts, such as ensuring equitable treatment between large corporations and SMEs in credit approval processes. Thus, applying a single quantitative fairness definition as a universal remedy across all sectors is impractical. This reasoning underpins our reliance on the causal framework, previous studies \citep{gastwirth1997statistical, pfohl_counterfactual_2019, kim_counterfactual_2021, kusner_counterfactual_2017, chiappa_path-specific_2019, imai2023experimental, imai2013experimental} have demonstrated the efficacy of causal inference techniques, with counterfactual causal inference standing out as particularly prominent. Counterfactual reasoning critically examines and establishes causal connections by contemplating hypothetical scenarios under altered conditions (e.g., “If the applicant were not a female, would the application be approved for a loan?”). While counterfactual approaches to fairness have been previously suggested \citep{kusner_counterfactual_2017}, a key limitation of counterfactual fairness is the unidentifiability from observational data \citep{wu_counterfactual_2019}. To address this, our framework proves that individuals or organisations with similar joint distributions could be identified as counterfactual instances. We identify these counterfactual instances as peers. We are motivated by the peer comparison perception \citep{li2016behavior}, which involves the differential treatment experienced by individuals compared to their peers. When an individual’s treatment is consistent with that of their peer group, perceptions of bias tend to diminish. 

Building on this insight, we introduce a novel concept termed ``\textit{peer-induced fairness}", which leverages the strengths of counterfactual fairness while addressing its limitations, thereby creating a more reliable and robust tool for auditing algorithmic fairness. 
This framework also serves as a valuable self-assessment tool, which is increasingly crucial for the AI community in the development of products that must comply with the EU AI Act. 
Beyond its general contributions to the field of algorithmic fairness auditing and self-assessment, our paper has the following particular contributions: 
\textbf{First}, to the best of our knowledge, our framework is the first to formalise a practical concept of ``\textit{peer-induced fairness}'' specifically designed to audit algorithmic biases. This comprehensive framework goes beyond the initial stage of basic detection, enabling users to evaluate externally obtained data without accessing the decision-making process or the underlying algorithmic model.
\textbf{Second}, our framework is universal and versatile in handling different levels of data quality, including datasets with highly skewed distributions of protected attributes issues often overlooked or inadequately addressed in previous studies. 
\textbf{Third}, when it is considered as a self-assessment tool, the ``peer-induced fairness" framework not only provides conclusions from self-assessment but also offers insightful explanations through peer comparisons, enhancing transparency and explainability.
\textbf{Fourth}, we demonstrate the practical utility of the ``\textit{peer-induced fairness}'' framework in uncovering algorithmic fairness issues related to small and medium-sized enterprises (SMEs) access to finance, particularly in scenarios where AI systems are used for decision-making. To the best of our knowledge, this is the first study of algorithmic bias on SMEs' access to finance\footnote{\cite{lu_cohort_2023} proposed a method for assessing the discrimination in ground truth $Y$ in SMEs' access to finance, rather than in the algorithmic predictions (\(\hat{Y}\)) made by AI systems}. It also highlights the effectiveness of our framework as a self-assessment tool in real-world applications. Besides, our framework is universal, applicable across different domains and capable of addressing various types of protected attributes. For example, it expands the protected attribute from the customer level (e.g., gender and race) to the organisation level (i.e., firm size).

The remainder of this paper is organised as follows. Section~\ref{sec:counterfactual-fairness} begins with an overview of the foundational concepts of counterfactual fairness and causal framework, including the representation of Single World Intervention Graphs (SWIGs). In Section~\ref{sec:peer-induced-fairness}, we introduce our ``peer-induced fairness'' framework in a step-by-step manner. Sections~\ref{sec:experiment} and~\ref{sec:results} detail our experimental procedures and present the empirical results, using the example of SMEs' access to finance in the UK. Finally, concluding remarks and further discussions are provided in Section~\ref{sec:conclusion}.

\section{Counterfactual fairness and its representation}
\label{sec:counterfactual-fairness}

Before presenting our framework, it is essential to introduce some key concepts related to counterfactual fairness and its representation. Various forms of counterfactual fairness have been proposed in the academic literature \citep{pfohl_counterfactual_2019, kim_counterfactual_2021, kusner_counterfactual_2017, wu_counterfactual_2019}. In this paper, we adopt the general framework outlined by \citet{wu_counterfactual_2019}.

Let \(S\) represent the set of protected features of an individual, which by definition, must not be subject to bias under any fairness doctrine. Additionally, let \(\bm{Z}\) represent the set of unprotected features, with \(\bm{X} \subseteq \bm{Z}\) specifying the subset of \textit{observable} features for any given individual. The outcome of the decision-making process, potentially influenced by historical biases, is denoted by \(Y\). We utilise a historical dataset \(\mathcal{D}\), sampled from a distribution \(\mathbb{P}(\bm{Z}, S, Y)\), to train a classifier \(f: (\bm{Z}, S) \mapsto \hat{Y}\), where \(\hat{Y}\) is the prediction generated by a machine learning algorithm aiming to estimate \(Y\). The causal structure underlying the distribution \(\mathbb{P}(\bm{Z}, S, \hat{Y})\) is represented by a graph causal model \(\mathcal{G}\).

\begin{definition}[Counterfactual fairness]\label{def:counterfactual-fairness}
Given a set of features $\bm{X} \subseteq \bm{Z}$, a classifier $f: (\bm{X}, S) \mapsto \hat{Y}$ is counterfactually fair with respect to $\bm{X}$ if under any observable context $\bm{X} = \bm{x}$ and $S = s$, 
    \begin{equation}
        \PP(\hat{Y}_{S \leftarrow s} = y | \bm{X} = \bm{x}, S = s) = \PP(\hat{Y}_{S \leftarrow s'} = y |  \bm{X} = \bm{x}, S = s),
    \end{equation}
for all $y$ and for any value $s'$ attainable by $S$.
\end{definition}

For a binary protected feature and a dichotomous decision outcome, a simplified version can be formulated.
\begin{definition}\label{def:counterfactual-fairness2}
Given a set of features $\bm{X} \subseteq \bm{Z}$, a binary classifier $f: (\bm{X}, S) \mapsto \hat{Y}$ is counterfactually fair with respect to $\bm{X}$ if under any observable context $\bm{X} = \bm{x}$ and $S = s_{-}$, 
    \begin{equation}
        \PP(\hat{Y}_{S \leftarrow s_{-}} = 1 | \bm{X} = \bm{x}, S = s_{-}) = \PP(\hat{Y}_{S \leftarrow s_{+}} = 1 | \bm{X} = \bm{x}, S = s_{-}),
    \end{equation}
for all $y$ and $S=\{s_{+}, s_{-}\}$.
\end{definition}

For illustrative purposes, imagine a scenario - loan applications using a predictive model, which determines the decision outcome, represented as \(\hat{Y}\). 
Let us focus on an application by a female, denoted by \(s_{-}\) with a specific profile \(\bm{x}\). The likelihood that this applicant received a favourable outcome is expressed as \(\PP(\hat{Y}|s_{-}, \bm{x})\), which is equivalent to \(\PP(\hat{Y}_{S \leftarrow s_{-}}=1|S = s_{-}, \bm{X} = \bm{x})\) by maintaining the protected feature (i.e., gender) unaltered. Suppose, hypothetically, that this applicant's protected feature is changed from \(s_{-}\) to \(s_{+}\). The probability of a favourable outcome after such a counterfactual modification is denoted by \(\PP(\hat{Y}_{s_{+}}|s_{-}, \bm{x})\). Counterfactual fairness is achieved when the probabilities \(\PP(\hat{Y}_{s_{-}}|s_{-}, \bm{x})\) and \(\PP(\hat{Y}_{s_{+}}|s_{-}, \bm{x})\) are equal, suggesting that the treatment of the applicant would remain consistent irrespective of their membership. This condition underscores the essence of counterfactual fairness, where the decision-making process is indifferent to changes in the protected features.

\begin{figure}[htbp]
    \centering
    \begin{subfigure}[b]{0.49\textwidth}
        \centering
        \begin{tikzpicture}
            \tikzset{line width=1pt, outer sep=0pt,
            ell/.style={draw,fill=white, inner sep=2pt,
            line width=1pt},
            swig vsplit={gap=3pt,
            line color right=red}};
            \node[name=a1,shape=swig vsplit]{
                \nodepart{left}{$S$}
                \nodepart[red]{right}{$s_{-}$} };
            \node[name=a2,shape=swig vsplit,
            above right=5mm of a1]{
                \nodepart{left}{$\bm{X}({\color{red}s_{-}})$}
                \nodepart[red]{right}{$\bm{x}$} };
            \node[name=y, below right=5mm of a2, ell, shape=ellipse]{$Y({\color{red}s_{-}},{\color{red}\bm{x}})$};
            \draw[blue, ->,line width=1pt, >=stealth]
            (a1) edge (a2)
            (a1) edge (y)
            (a2) edge (y);
        \end{tikzpicture}
        \caption{Actual Scenario: $\mathcal{G}(s_{-}, \bm{x})$}
    \end{subfigure}
    \hfill
    \begin{subfigure}[b]{0.49\textwidth}
        \centering
        \begin{tikzpicture}
            \tikzset{line width=1pt, outer sep=0pt,
            ell/.style={draw,fill=white, inner sep=2pt,
            line width=1pt},
            swig vsplit={gap=3pt,
            line color right=red}};
            \node[name=a1,shape=swig vsplit]{
                \nodepart{left}{$S$}
                \nodepart[red]{right}{$s_{+}$} };
            \node[name=a2,shape=swig vsplit,
            above right=5mm of a1]{
                \nodepart{left}{$\bm{X}({\color{red}s_{-}})$}
                \nodepart[red]{right}{$\bm{x}$} };
            \node[name=y, below right=5mm of a2, ell, shape=ellipse]{$Y({\color{red}s_{+}},{\color{red}\bm{x}})$};
            \draw[blue, ->,line width=1pt, >=stealth]
            (a1) edge (y)
            (a2) edge (y);
            \draw[green,dashed, ->,line width=1pt, >=stealth]
            (a1) edge (a2);
        \end{tikzpicture}
        \caption{Counterfactual Scenario: $\mathcal{\tilde{G}}(s_{+}, \bm{x})$}
    \end{subfigure}
        \hfill
    \begin{subfigure}[b]{0.49\textwidth}
        \centering
        \begin{tikzpicture}
            \tikzset{line width=1pt, outer sep=0pt,
            ell/.style={draw,fill=white, inner sep=2pt,
            line width=1pt},
            swig vsplit={gap=3pt,
            line color right=red}};
            \node[name=a1,shape=swig vsplit]{
                \nodepart{left}{$S$}
                \nodepart[red]{right}{$s_{+}$} };
            \node[name=a2,shape=swig vsplit,
            above right=5mm of a1]{
                \nodepart{left}{$\bm{X}({\color{red}s_{+}})$}
                \nodepart[red]{right}{$\bm{x}'$} };
            \node[name=y, below right=5mm of a2, ell, shape=ellipse]{$Y({\color{red}s_{+}},{\color{red}\bm{x}'})$};
            \draw[blue, ->,line width=1pt, >=stealth]
            (a1) edge (y)
            (a2) edge (y)
            (a1) edge (a2);
        \end{tikzpicture}
        \caption{Actual Scenario: $\mathcal{G}(s_{+}, \bm{x}')$}
    \end{subfigure}
    \caption{\textbf{SWIGs for Graphical Causal Models (GCM)}.
    The nodes with black border represent random variables, while red ones indicate fixed values of random variables, representing experimental interventions. Arrows depict causal relationships between variables.
    \textbf{(a)}: The SWIG $\mathcal{G}(s_{-}, \bm{x})$ represents the actual scenario for an individual with features $(s_{-}, \bm{x})$. 
    \textbf{(b)}: The SWIG $\mathcal{\tilde{G}}(s_{+}, \bm{x})$ illustrates the counterfactual scenario, assuming the individual's protected feature changes from $s_{-}$ to $s_{+}$, while their other features $\bm{x}$ remain the same.
    \textbf{(c)}: The SWIG $\mathcal{G}(s_{+}, \bm{x}')$ represents the actual scenario for an individual with features $(s_{+}, \bm{x}')$.
    The actual SWIG $\mathcal{G}(s_{-}, \bm{x})$ corresponds to the conditional distribution $\hat{Y}_{s_{-}} | s_{-}, \bm{x}$. Conversely, in the counterfactual SWIG $\mathcal{\tilde{G}}(s_{+}, \bm{x})$ refers to $\hat{Y}_{s_{+}} | s_{-}, \bm{x}$, denoting the outcome distribution had the individual been featured with $s_{+}$, given that the actual features are $(s_{-}, \bm{x})$. Thus the directed link from $s_{+}$ to $\bm{X}(s_{-})$ is not the fact (shown in green colour).
    Note: $\mathcal{\tilde{G}}(s_{+},\bm{x}) \neq \mathcal{G}(s_{+},\bm{x}')$ because $\mathcal{\tilde{G}}(s_{+},\bm{x})$ is counterfactual scenario with actual features $(s_{-},\bm{x})$ and $\mathcal{G}(s_{+},\bm{x}')$ is the fact with features $(s_{+},\bm{x}')$.
    }
    \label{fig:swig-simple}
\end{figure}
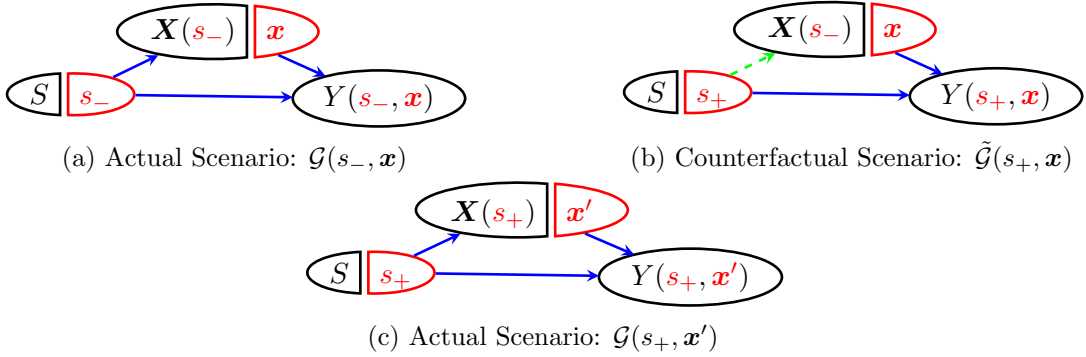

A more nuanced comprehension of counterfactual fairness may be facilitated through the lens of SWIGs \citep{richardson_2013_single}. Consider an individual belonging to a disadvantaged group \( s_{-} \), characterised by features \( \bm{x} \). The label \( s_{-} \) could exert a direct influence on the outcome \( Y \), or it may indirectly impact \( Y \) through its effect on other observable features \( \bm{X} \). If we postulate a counterfactual scenario in which the individual's group designation changes from \( s_{-} \) to \( s_{+} \), the corresponding Graphical Causal Models (GCMs) for both actual and hypothetical situations can be depicted using SWIGs, as illustrated in \autoref{fig:swig-simple}. Counterfactual fairness is attained if the predictor, consistent with the actual GCM and the counterfactual GCM, yields identical probabilities for the outcome given the specific features \( (s_{-}, \bm{x}) \).

Next, let us review some pivotal conclusions derived from the SWIGs as depicted in \autoref{fig:swig-simple} (a) and propose some notations. A key aspect we will discuss is the factorisation properties of the joint distribution of all variables within a SWIG, applicable to any protected feature \(s_{-}, s_{+}\) and other features \(\bm{x}\), which can be mathematically represented as follows:
\begin{equation}
    \mathcal{G}(s, \bm{x}): \mathbb{P}(S, \bm{X}(s), Y(s, \bm{x}) ) = \mathbb{P}(S) \cdot \mathbb{P}(\bm{X}(s)) \cdot \mathbb{P}(Y(s, \bm{x})), s \in \{s_{-}, s_{+}\}. \label{eq:G(s, x)}
\end{equation}
Furthermore, the modularity property is observed where:
\begin{align}
    & \mathbb{P}(\bm{X}(s) = \bm{x}) = \mathbb{P}(\bm{X}=\bm{x} | S = s), s \in \{s_{-}, s_{+}\}, \label{eq:X(s)} \\
    & \PP(Y(s, \bm{x}) = y) = \PP(Y=y | \bm{X}=\bm{x}, S=s), s \in \{s_{-}, s_{+}\}, \label{eq:Y(s,x)}
\end{align}
highlighting the left-hand side is the potential outcome while the right-hand side is the observational conditional probability.
In the context of the counterfactual scenario with actual features $(s_{-}, \bm{x})$ shown in \autoref{fig:swig-simple} (b), a similar joint distribution is applicable:
\begin{equation}
    \mathcal{\tilde{G}}(s_{+}, \bm{x}): \mathbb{P}(S, \bm{X}(s_{-}), Y(s_{+}, \bm{x}) ) = \mathbb{P}(S) \cdot \mathbb{P}(\bm{X}(s_{-})) \cdot \mathbb{P}(Y(s_{+}, \bm{x})). \label{eq:CG(s+, x)}
\end{equation} 

While the above concept of counterfactual fairness is theoretically straightforward and can be easily described, its application in practice is hampered by the challenges in identifying counterfactual outcomes from observational data in certain scenarios, as highlighted by \citet{wu_counterfactual_2019}. Specifically, the probability \(\mathbb{P}(\hat{Y}_{s_{+}}|s_{-}, \bm{x})\) as a potential outcome remains elusive for direct calculation due to its unidentifiability. This creates a significant challenge for regulators, making it difficult to implement algorithmic bias auditing.

\section{Peer-induced fairness with causal method}\label{sec:peer-induced-fairness}
In this section, we propose a practical approximation method that utilises peer comparison as an effective strategy, in order to navigate the above impediment and facilitate a feasible implementation of counterfactual fairness. We then introduce our ``peer-induced fairness'' framework and algorithmic bias-detecting methods after providing the peer identification concept. 

\subsection{Discrimination from peer comparisons}

The phenomenon of discrimination, a ubiquitous aspect of daily life, is extensively explored within cognitive science literature. Research indicates that perceptions of discrimination are shaped not only by personal experiences but also through comparisons with peers who, despite possessing similar capabilities, skills, or knowledge, experience differential treatment, leading to missed opportunities. These perceptions are cultivated both through individual encounters and the lens of peer experiences \citep{li2016behavior}. When an individual's treatment aligns with that of their peer group, perceptions of being biased tend to diminish. Studies have shown that social and financial ties are more likely to form among individuals who share similarities in revenue levels, consumption behaviours, educational background, class, gender, race, or creditworthiness, illustrating a preference for homogeneity \citep{li_credit_2020, haenlein_social_2011, goel_predicting_2014, wei_credit_2016}. 

The insights from cognitive science highlight the importance of understanding how comparative experiences shape perceptions of discrimination among individuals and organisations. These groups may not only perceive but also actually experience biased outcomes when compared to their peers. Such perceived or real biases could erode trust in automated systems and, more broadly, undermine confidence in the regulators intended to ensure fairness. To mitigate these risks, regulators need to ensure that auditing frameworks are sensitive to these subtler forms of discrimination, which can arise from differences in treatment relative to peers.

\subsection{Fairness through peer observations}\label{sec:fairness-through-peers}
Building on the concept of bias through peer comparisons discussed previously, we propose a more rigorous mathematical representation to demonstrate this idea effectively.

Consider an individual \(A\) from a protected group with a protected status \(S = s_{-}\) and other unprotected features \(\bm{X} = \bm{x}\), denoted as \(A = (s_{-}, \bm{x})\). Assuming the protected and unprotected groups are comparable, if there exists a group of peers $\mathcal{C} = \{C_1, C_2, \cdots\}$ from the unprotected group \(S = s_{+}\), represented as \( \{(s_{+}, \bm{x}_1), (s_{+}, \bm{x}_2), \cdots\} \), forming an $A$-oriented network. 
We use the expectation of the probability \(\mathbb{P}(\hat{Y}_{s_{+}}| s_{+}, \bm{x}_j)\) across these peers  $\mathcal{C}$ to approximate the counterfactual \(\mathbb{P}(\hat{Y}_{s_{+}}| s_{-}, \bm{x})\), mathematically expressed as:
\begin{equation}\label{eq:peer-approximation}
    \mathbb{P}(\hat{Y}_{s_{+}}| s_{-}, \bm{x}) \approx \mathbb{E}_{(s_{+}, \bm{x}_j) \in \mathcal{C}} [\mathbb{P}(\hat{Y}_{s_{+}}| s_{+}, \bm{x}_j)],
\end{equation}
where $\mathbb{E}{[\cdot]}$ is the expectation (or average) notation. 
This peer-based counterfactual approximation is intuitive, adhering to the non-discrimination principle where, ideally, the unobserved counterfactual probability aligns consistently with the average observed among peers. The method avoids the necessity for conventional statistical estimations within the protected group by employing resilient counterfactual statistics obtained from adequately represented peer groups. It adeptly addresses data scarcity within the protected group.

\subsection{Peer definition and identification}\label{sec:peer-identification}

While the counterfactual predictive probability is provided in Eq.~\eqref{eq:peer-approximation}, a key question remains: ``What is a peer'' in mathematical terms? A precise definition and identification of peers are crucial before initiating peer comparisons, as they ensure accurate and reliable assessments. This clarity is essential for regulators to effectively audit potential biases in algorithmic decision-making.

\begin{definition}[$\delta$-peer]\label{def:delta-peer}
Let an individual \(A\) belong to a protected group, characterised by a protected feature \(S = s_{-}\) and a set of unprotected features \(\bm{X} = \bm{x}_0\), represented as \(A = (s_{-}, \bm{x}_0)\). 
Assuming there exists a set of individuals \( \mathcal{B} =\{B_1, B_2, \cdots \}\) from the unprotected group, where \(B_i = (s_{+}, \bm{x}_i)\) for \(i = 1, 2, \ldots \). 
An individual \(C \in \mathcal{B}\) is defined as \(\delta\)-peer of \(A\) if the difference in joint distributions between \(C\)'s actual SWIG, \(\mathcal{G}(s_{+}, \bm{x}_j)\), and \(A\)'s counterfactual SWIG, \(\mathcal{\tilde{G}}(s_{+}, \bm{x}_0)\), is less than a threshold \(\delta\), 
    \begin{equation}
        \left| \mathbb{P}(\mathcal{G}(s_{+}, \bm{x}_j)) - \mathbb{P}(\mathcal{\tilde{G}}(s_{+}, \bm{x}_0)) \right| < \delta,
    \end{equation}
where \(\mathbb{P}(\mathcal{G}(s_{+}, \bm{x}_j)) = \mathbb{P}(S, \bm{X}(s_{+}), Y(s_{+}, \bm{x}_j))\) and \(\mathbb{P}(\mathcal{\tilde{G}}(s_{+}, \bm{x}_0)) = \mathbb{P}(S, \bm{X}(s_{-}), Y(s_{+}, \bm{x}_0))\).
\end{definition}

In a graphical causal model, the concept of a peer is defined through the interrelations among three random variables: \(S\), \(\bm{X}\), and \(Y\). To enable rigorous and unbiased comparisons, it is crucial that a peer exhibits a joint distribution similar to the counterfactual scenario. However, since the counterfactual scenario is inherently unobservable, it is rarely possible to observe the exact same \(\bm{X}\) and \(Y\) in another group defined by different protected attributes.

Although we have the mathematical representation of a \(\delta\)-peer in Definition~\ref{def:delta-peer}, its practical implementation faces significant challenges. A major obstacle is the difficulty in calculating \(\mathbb{P}(Y(s_{+}, \bm{x}))\) from Eq.~\eqref{eq:CG(s+, x)} for the counterfactual scenario \((s_{+}, \bm{x})\), which is essential for evaluating peer similarity. This difficulty arises because \(\bm{x}\) represents the unprotected features for the protected group, where the direct calculation of this probability is often infeasible due to the lack of observational data.

To address this and develop a more feasible approach for peer selection, we re-examine Eq.~\eqref{eq:G(s, x)} and Eq.~\eqref{eq:CG(s+, x)}. Since it is not feasible to directly derive \(\mathcal{\tilde{G}}(s_{+}, \bm{x})\) from observational data, we have no choice but use the information from \(\mathcal{G}(s_{+}, \bm{x})\) as a proxy for approximation, which has been discussed in Section~\ref{sec:fairness-through-peers}. Upon comparing Eq.~\eqref{eq:G(s, x)} and Eq.~\eqref{eq:CG(s+, x)}, the difference lies in the terms $\bm{X}$ and $Y$. 
Referring to \autoref{fig:swig-simple} (a) and considering \(\bm{x}_0\) as the observable unprotected features of an individual from the protected group \(S = s_{-}\), we can compute \(\mathbb{P}(\bm{X}(s_{-}) = \bm{x}_0)\) using Bayes' formula:
\begin{align}
    \mathbb{P}(\bm{X}(s_{-}) = \bm{x}_0) &= \mathbb{P}(\bm{X}=\bm{x}_0 | S = s_{-}) \nonumber \\ 
    &= \frac{\mathbb{P}(\bm{X} = \bm{x}_0) \mathbb{P}(S = s_{-}|\bm{X}=\bm{x}_0)}{\mathbb{P}(S = s_{-})}. \label{eq:ps-X(s-)}
\end{align}
Similarly, we can determine \(\mathbb{P}(\bm{X}(s_{+}) = \bm{x}_0)\):
\begin{align}
    \mathbb{P}(\bm{X}(s_{+}) = \bm{x}_0) &= \mathbb{P}(\bm{X}=\bm{x}_0 | S = s_{+}) \nonumber \\ 
    &= \frac{\mathbb{P}(\bm{X} = \bm{x}_0) \mathbb{P}(S = s_{+}|\bm{X}=\bm{x}_0)}{\mathbb{P}(S = s_{+})}. \label{eq:ps-X(s+)}
\end{align}
However, because \(\bm{x}_0\) are the observable unprotected features for an individual from the protected group \(S = s_{-}\), estimating \(\mathbb{P}(S = s_{+}|\bm{X}=\bm{x}_0)\) directly is not feasible. Given that \(S\) represents a binary set, we can infer $\mathbb{P}(S = s_{+}|\bm{X}=\bm{x}_0) = 1 - \mathbb{P}(S = s_{-}|\bm{X}=\bm{x}_0)$.
We can rewrite \(\mathbb{P}(\bm{X}(s_{-}))\) and \(\mathbb{P}(\bm{X}(s_{+}))\) in a unified representation,
\begin{equation}
    \PP(\bm{X}(s) = \bm{x}) = \mathbb{P}(\bm{X} = \bm{x}) \xi(s, \bm{x}), \label{eq:P(X) = x}
\end{equation}
where \(\xi(s, \bm{x})\) is defined as the \emph{identification coefficient (IC)}. This coefficient adjusts the probability values to reflect the conditions of being either a factual or counterfactual group, and is given by:
\begin{equation}
    \xi(s, \bm{x}) = \begin{cases} 
    \frac{1}{\mathbb{P}(S = s_{-})} \cdot \mathbb{P}(S = s_{-}|\bm{X}=\bm{x}), & \text{if } s=s_{-}, \\
    \frac{1}{1 - \mathbb{P}(S = s_{-})} \cdot (1 - \mathbb{P}(S = s_{-}|\bm{X}=\bm{x})), &  \text{if } s=s_+.
    \end{cases}
\end{equation}

Although direct evaluation of the joint distribution \(\mathcal{\tilde{G}}(s_{+}, \bm{x})\) is not feasible, we can facilitate the comparison by utilising the computable \(\xi(s, \bm{x})\). 
This approach hinges on quantitative comparison and addresses the critical question: ``\textit{How can peers be identified}?''. Traditional methods often employ multi-dimensional matching to identify similar individuals within datasets, typically focusing on unprotected features \(\bm{X}\). However, the causal impact of protected features \(S\) on \(\bm{X}\), coupled with the high dimensionality of \(\bm{X}\), poses significant challenges to the efficacy of these conventional matching techniques. The complexity introduced by the curse of dimensionality makes the straightforward application of these methods problematic. 

We propose a practical approach to implement a \(\delta\)-peer identification algorithm. The approach utilises information from the counterpart group, effectively addressing the issues of data scarcity and imbalance theoretically.

\begin{theorem}[$\delta$-peer identification]\label{thm:delta-peer-identification}
Consider an individual $A = (s_{-}, \bm{x}_0)$ and assuming there are a group of individuals $\mathcal{B} = \{B_1, B_2, \cdots$ \} from unprotected group, where $B_j = (s_{+}, \bm{x}_j)$. An individual $C \in \mathcal{B}$ is identified as a $\delta$-peer of $A$ if:
        \begin{equation}
            |\xi(s_{-}, \bm{x}_0) - \xi(s_{+}, \bm{x}_j)| < \delta.
        \end{equation}
\end{theorem}

Theorem~\ref{thm:delta-peer-identification} provides a sufficient condition for Definition~\ref{def:delta-peer}, with the proof detailed in \ref{sec:appendix_proof_thm}. Based on this, we propose using the \emph{IC} for peer identification as a practical alternative to the infeasible joint distribution. By enhancing the identification of suitable peers, regulators can effectively audit potential biases in decision-making systems in real-world scenarios using feasible methods. More practically, we can also implement the idea as an algorithm shown in \ref{sec:appendix_algo} to identify all peers in the dataset step by step. This algorithm, by applying a similarity threshold \(\delta\), is grounded in cognitive science perception of discrimination, ensuring that peers are selected for meaningful comparison based on their \emph{IC} similarities to the protected individual.

\subsection{Peer-induced fairness}
\label{sec:Peer-induced fairness}

Following the idea of peer comparison, definition, and identification, we can now introduce the concept of ``peer-induced fairness''. 
\begin{definition}[$(\delta, f)$-peer-induced fairness\footnote{Although the term ``peer-induced fairness'' has been used in other contexts, as noted by \citep{ho2009peer, li2016behavior}, our work is distinct.}]\label{def:peer-induced-fairness}
Consider an individual $A=(s_{-}, \bm{x}_0)$ and assuming $A$ has a number of $\delta$-peers $\mathcal{C} = \{C_1, C_2, \cdots\}$ where $C_j = (s_{+}, \bm{x}_j)$.
$A$ is said  to be \emph{fairly treated} by the peers subject to $(\delta, f)$ if and only if 
    \begin{equation}
       \mathbb{P}(\hat{Y}_{s_{-}}| s_{-}, \bm{x}_0) = \mathbb{E}_{C_j \in \mathcal{C}} [\mathbb{P}(\hat{Y}_{s_{+}}| C_j)] , \label{eq:peer-induced-fairness}
    \end{equation}
where $\hat{Y}$ is the predictive outcome provided with the classifier $f$.
\end{definition}

As discussed in previous sections, while we can directly estimate \(\mathbb{P}(\hat{Y}_{s_{-}}|s_{-}, \bm{x})\) from individual observations, estimating the expected value \(\mathbb{E}_{C_j \in \mathcal{C}} [\mathbb{P}(\hat{Y}_{s_{+}}| C_j)]\) presents challenges due to the limited number of observations available for $\delta$-peers. 
Consequently, we have to rely on observable peers to approximate the population mean. To formalise this, we introduce the random variable
\begin{equation}
    T_j = \mathbb{P}(\hat{Y}_{s_{+}}| C_j).
\end{equation}
Upon examining the distribution of \(T_j\), we find that it does not always follow a normal distribution, with details presented in Supplementary Materials. Therefore, we randomly select a subset of peers and use the sample mean to estimate the population mean,
\begin{equation}
    \bar{T} = \frac{1}{K} \sum_{j=1}^K T_j, \label{eq:sample-mean}
\end{equation}
where \(K\) is a large enough number of peers in the subset. 

According to the Central Limit Theorem, the sample mean $\bar{T}$ follows a normal distribution, and thus \(\mathbb{E}[\bar{T}]\), can be employed to estimate the overall predictive probabilities of favourable outcomes among peers, denoted as \(\mathbb{E}[T] = \mu\). Based on this, we propose a proposition that a synthetic individual, defined using \emph{IC} \footnote{Although the synthetic individual is defined by \emph{IC}, the corresponding predictive favourable outcome probabilities calculation should follow Eq.~\eqref{eq:sample-mean}.}, can also be considered as a \(\delta\)-peer (The proof is given in \ref{sec:appendix_proof_prop}).

\begin{proposition}\label{prop:synthetic}
    Let \(A\) be an individual and \(\mathcal{C} = \{C_1, C_2, \ldots\}\) denote all of \(A\)'s \(\delta\)-peers. Define a synthetic individual \(\bar{T}_i\) using the average \emph{IC} of any subset \(\mathcal{C}_i\) of $K$ peers, where \(\mathcal{C}_i = \{C_1^i, \ldots, C_K^i\} \subseteq \mathcal{C}\), $i\in\{1,2, \ldots, N\}$ and \(C_j^i\) represents the \(j\)-th peer in the \(i\)-th selection with the unprotected feature \(\bm{x}_j^i\). This synthetic individual \(\bar{T}_i =\sum_{j=1}^K \mathbb{P}(\hat{Y}_{s_{+}}|C_{j}^i) / K \) can also be considered as a \(\delta\)-peer of \(A\).
\end{proposition}

Consequently, by randomly selecting \(K\) peers from the set of all observed \(\delta\)-peers \(N\) times, we compute the predictive favourable outcome probabilities \(\bar{T}_i\) for each \(i\)-th selection. We then use the mean of the resulting sample mean distribution, \(\{\bar{T}_i\}_{i=1}^N\), consisting of all confirmed \(\delta\)-peers as per Proposition~\ref{prop:synthetic}, to estimate the overall mean \(\mu\) of favourable outcome probabilities across \emph{all} peers.

\subsection{Peer-induced fairness auditing}
\label{sec:hypothesis_testing}
Finally, to formalise the process of auditing whether an individual in a protected group is subjected to algorithmic bias, we take advantage of hypothesis testing. This framework is predicated on an appropriate threshold for peer identification \(\delta\) and a specific classifier $f$. It aims to test whether the sample mean distribution \(\{\bar{T}_i\}_{i=1}^N\) is statistically equivalent to \(\mathbb{P}(\hat{Y}_{s_{-}}| s_{-}, \bm{x})\). 
Since $\bar{T}_i$ follows a normal distribution and $N$ is a large enough number, our hypothesis is consistent with the standard $z$-test, which is designed to evaluate the presence of algorithmic bias statistically.

\begin{itemize}
    \item \textbf{$H_0$ (Null Hypothesis)}: The individual $A=(s_{-}, \bm{x}_0)$ is equally treated according to $(\delta, f)$-``peer-induced fairness'' criterion,
    \begin{equation}
        H_0: \mathbb{E}[\bar{T}_i] = \mathbb{P}(\hat{Y}_{s_{-}}| s_{-}, \bm{x}_0).
    \end{equation}
    
    \item \textbf{$H_1$ (Alternative Hypothesis)}: 
    The individual \(A\) is subject to algorithmic bias under $(\delta, f)$-``peer-induced fairness'' criterion, which is evidenced by a significant disparity in treatment compared to their unprotected peers,

    \begin{equation}
        H_1: \mathbb{E}[\bar{T}_i] \neq \mathbb{P}(\hat{Y}_{s_{-}}| s_{-}, \bm{x}_0).
    \end{equation}
\end{itemize}

Furthermore, it is also potential to consider two additional scenarios with one-sided tests: checking whether the individual is algorithmically discriminated against, where \( H_2: \mathbb{E}[\bar{T}_i] < \mathbb{P}(\hat{Y}_{s_{-}}| s_{-}, \bm{x}_0) \), or algorithmically benefited, where \( H_3: \mathbb{E}[\bar{T}_i] > \mathbb{P}(\hat{Y}_{s_{-}}| s_{-}, \bm{x}_0) \).

\subsection{Overall auditing workflow }
\label{sec:framework}
\begin{figure}[H]
\begin{tikzpicture}[node distance=1.5cm]
\tikzstyle{startstop} = [rectangle, rounded corners, minimum width=3cm, minimum height=1cm, text centered, draw=black, fill=red!30]
\tikzstyle{process} = [rectangle, minimum width=3cm, minimum height=1cm, text centered, draw=black, fill=orange!30]
\tikzstyle{decision} = [diamond, minimum width=3cm, minimum height=1cm, text centered, draw=black, fill=green!30]
\tikzstyle{arrow} = [thick,->,>=stealth]
\tikzstyle{groupbox} = [draw, dashed, inner sep=0.2cm, rounded corners, label={[anchor=south]north:Steps for Each Observation $a \in A $}] 
\tikzstyle{samplebox} = [draw, dashed, inner sep=0.2cm, rounded corners, label={[xshift=-1.5cm]north:Sampling $N$ times}] 

\node (start) [startstop] {Start: AI decision system \((x, s, y)\)};
\node (split) [decision, below of=start, yshift=-0.5cm] {Split by \(s\)};
\node (groupA) [process, below left of=split, xshift=-3.0cm] {Protected group A (\(s = s\_\))};
\node (groupB) [process, below right of=split, xshift=3.0cm] {Non-protected group B (\(s = s_{+}\))};
\node (task1) [process, below left of=groupA, yshift=-5.0cm, xshift=1.5cm] {Compute \(\PP(\hat{Y} = 1|s_, x)\) };
\node (task2) [process, below of=groupA, yshift=-2.2cm] {Compute $IC$};
\node (task3) [process, below of=groupB, yshift=-0.5cm] {Compute $IC$};
\node (peers) [process, below of=task3, yshift=-0.2cm] {Identify peers in B using $IC$, denote $\mathcal{C}(a)$};
\node (sample) [process, below of=peers, yshift=-0.5cm] {Choose \(K\) samples from $\mathcal{C}(a)$};
\node (expectation) [process, below of=sample, yshift=-0.0cm] {Compute \(\mathbb{E}[\bar{T}_a]\)};
\node (hypothesis_results) [decision, below of=split, yshift=-6.9cm] {Hypothesis test};
\node (discriminated) [process, below left of=hypothesis_results, xshift=-2.5cm, yshift=-0.5cm] {Discriminated};
\node (fairly) [process, below of=hypothesis_results, yshift=-1cm] {Fairly treated};
\node (privileged) [process, below right of=hypothesis_results, xshift=2.5cm, yshift=-0.5cm] {Privileged};
\node (check) [decision, below of=fairly, yshift=-0.2cm] {\(y = 0\)?};
\node (explanation) [process, below of=check, yshift=-0.4cm] {Provide explanation};
\node (end) [startstop, below of=explanation] {End: Audit report};

\node (groupbox) [groupbox, fit={(task1) (task2) (task3) (peers) (sample) (expectation) (hypothesis_results) (fairly) (privileged) (check) (explanation) }] {};

\node (samplebox) [samplebox, fit={(sample)}] {};

\draw [arrow] (start) -- (split);
\draw [arrow] (split) -- node[anchor=east, pos=0.6, xshift=-0.1cm, yshift=0.2cm]  {\(s = s\_\)} (groupA);
\draw [arrow] (split) -- node[anchor=west, pos=0.6, xshift=0.1cm, yshift=0.2cm] {\(s = s_{+}\)} (groupB);
\draw [arrow] (groupA.west) -- ++(-0.5,0) |-(task1.west);
\draw [arrow] (groupA) --(task2);
\draw [arrow] (task2) -- (peers);
\draw [arrow] (groupB) -- (task3);
\draw [arrow] (task3) -- (peers);
\draw [arrow] (task1) |- (hypothesis_results);
\draw [arrow] (peers) -- (sample);
\draw [arrow] (sample) -- (expectation);
\draw [arrow] (expectation) -- (hypothesis_results);
\draw [arrow] (hypothesis_results) -- (fairly);
\draw [arrow] (hypothesis_results) -- (discriminated);
\draw [arrow] (hypothesis_results) -- (privileged);
\draw [arrow] (fairly) -- (check);
\draw [arrow] (check) -- node[anchor=east] {Yes} (explanation);
\draw [arrow] (explanation) -- (end);
\draw [arrow] (discriminated) |- (end);
\draw [arrow] (privileged) |- (end);

\end{tikzpicture}

  \caption{The overall auditing workflow. ``Compute \(\PP(\hat{Y} = 1|s_, x)\)'' step requires a given prediction model, ``Compute $IC$'' step requires a given fitting model, ``Find peers in B using $IC$, denote $\mathcal{C}(a)$'' step requires a given $\delta$. Although the fitting model is usually the same as the prediction model, distinct choices are also allowed.}
  \label{fig:framework}
\end{figure}
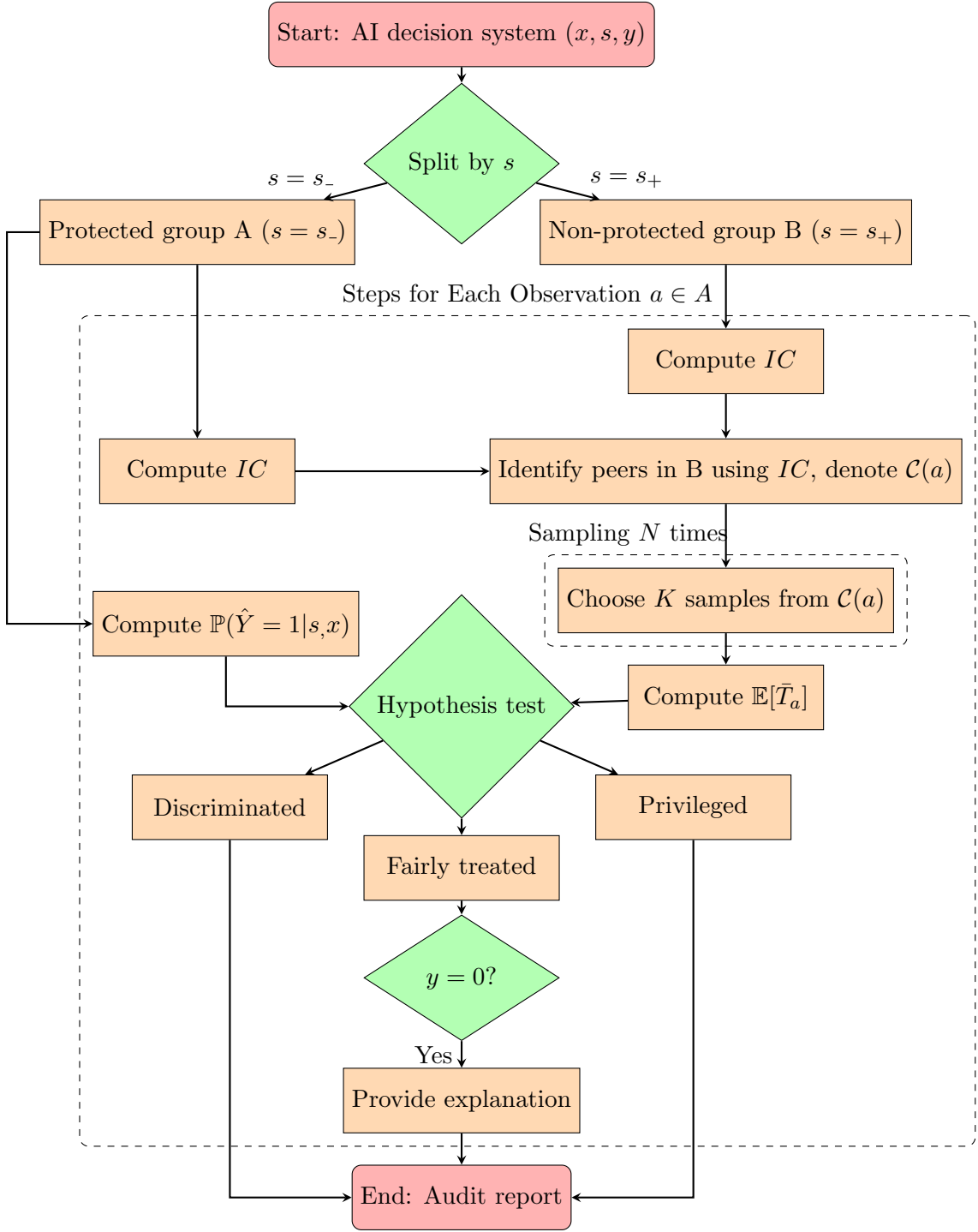

To illustrate the overall workflow of our ``peer-induced fairness'' tool, we present a flowchart in \autoref{fig:framework} that visualises the steps involved in assessing potential algorithmic bias in an AI decision system. As an auditing tool, this framework can effectively determine whether the outcomes produced by an AI decision system exhibit algorithmic bias against a particular protected group. The process is straightforward and can function as a plug-and-play tool for not only AI developers but also regulators without access to the underlying decision process.

Specifically, depending on the specific scenario, users can select different fitting and prediction models for computing $IC$ and predict \(\PP(\hat{Y} = 1|s_, x)\) respectively for each instance in the datasets. Moreover, to handle varying characteristics of protected attributes and different levels of data quality, users have the flexibility to adjust the threshold \(\delta\) in peer identification, which defines the degree of similarity required for an individual from a non-protected group to be considered a valid peer. This allows for a balance between the need for precise comparability and the practical constraints of the dataset.

A critical consideration in sampling is ensuring the use of the Central Limit Theorem, which typically requires at least 35 peers for each individual in the protected group and thus we can randomly select at least 30 from them. This should be taken into account when deciding on an appropriate threshold \(\delta\). A threshold that is too high may result in too many distinct peers, while one that is too low may lead to too few peers, preventing further analysis.

Additionally, the hypothesis testing of our framework is adaptable. Users can choose between one-sided or two-sided hypothesis tests, as discussed in Section~\ref{sec:hypothesis_testing}, and set significance levels based on their specific needs. For a two-sided hypothesis test, the framework categorises results into ``Fairly treated'' and ``Not fairly treated''. Advanced statistical tests can also be explored for more complex cases.

While there might be some misunderstandings and ambiguities in audit results, particularly when individuals who are treated fairly by the system still receive unfavourable decisions. Our framework addresses these issues by providing explanations. Explanations are offered only for fairly treated individuals. This allows offer actionable advice on how they can achieve more favourable outcomes, significantly enhancing the customer service experience. Additionally, this analysis helps identify specific areas relevant institutions should pay attention to, such as particular features where these individuals may be under-performing. Conversely, for those who experience discrimination or privilege, the problem lies within the AI decision system, not with the applicants. In such cases, efforts should be concentrated on improving the AI system rather than providing suggestions to the applicants.

Finally, as shown in \autoref{fig:framework}, the framework can also serve as a self-assessment tool before releasing an AI decision-making product. During internal testing, developers can easily integrate this framework to conduct various tests—such as using different prediction models, fitting models, and assessing different data qualities—even performing stress testing for algorithmic fairness.

\section{Experiment}
\label{sec:experiment}

To demonstrate the effectiveness of our proposed ``peer-induced fairness'' framework in auditing algorithmic fairness, and to examine the current state of credit scoring concerning fairness, we experiment on the SMEs' access to finance.

The dataset used for this study is collected from the UK Archive Small and Medium-Sized Enterprise Finance Monitor \citep{bdrc_continental_sme_2023}. The dataset compiles survey information on SMEs\footnote{SMEs included in this survey meet the four criteria: 1) employ no more than 250 individuals, 2) have an annual turnover not exceeding £25 million, 3) do not operate as social enterprises or non-profit organisations, and 4) are not owned by another company by more than 50\% \citep{bdrc_continental_sme_2023}.}, spanning from 2011Q1 to 2023Q4, with approximately 4,500 telephone interviews conducted per quarter across the UK. 
Each interview provides insights into the experiences of SMEs with external financing over the past 12 months, including their anticipated future financial needs and perceived obstacles to growth. It also details the characteristics of the SMEs and their owners or managers.

\begin{table}[H]
\renewcommand{\arraystretch}{0.8}
\small
\centering
\caption{Description and abbreviation of features, grouped by whether they are intrinsic. The final column presents the values alongside their corresponding percentages.}
\label{tab:description_dataset}
\begin{tabular}{>{\raggedright\arraybackslash}p{3.8cm}>{\raggedright\arraybackslash}p{2.5cm}>{\raggedright\arraybackslash}p{7.8cm}}
\toprule
\textbf{Features} & \textbf{Abbreviation} & \textbf{Value (Percentage)} \\
\multicolumn{3}{l}{\textbf{Non-intrinsic Features}} \\
previous turn-down & PT & no (90.94\%) \\
                   &    & yes (9.06\%) \\
                   \\
finance qualification & FQ & no (45.66\%) \\
                      &    & yes (54.34\%) \\
                      \\
written plan & WP & no (37.58\%) \\
             &    & yes (62.42\%) \\
             \\
risk & RI & minimal (19.59\%) \\
     &    & low (43.11\%) \\
     &    & average (25.98\%) \\
     &    & above average (11.31\%) \\
     \\
product/service development & PS & no (70.25\%) \\
                            &    & yes (29.75\%) \\
                            \\
business innovation & BI & no (40.16\%) \\
                    &    & yes (59.84\%) \\
                    \\
loss or profit & LP & loss (86.07\%) \\
               &    & broken even (8.69\%) \\
               &    & profit (5.25\%) \\
               \\
turnover growth rate & TG & grown more than 20\% (13.69\%) \\
                     &    & grown but by less than 20\% (40.33\%) \\
                     &    & stayed the same (33.69\%) \\
                     &    & declined (12.30\%) \\
                     \\
funds injection & FI & no (67.17\%) \\
                &    & yes (32.83\%) \\
                \\
credit purchase & CP & no (18.48\%) \\
                &    & yes (81.52\%) \\
                \\
regular management account & RM & no (19.06\%) \\
                           &    & yes (80.94\%) \\
\\
\multicolumn{3}{l}{\textbf{Intrinsic Features }} \\
principal & PR & construction (6.64\%) \\
          &    & agriculture, hunting and forestry (10.82\%) \\
          &    & fishing (12.01\%) \\
          &    & health and social work (12.62\%) \\
          &    & hotels and restaurants (11.68\%) \\
          &    & manufacturing (8.69\%) \\
          &    & real estate, renting and business activities (16.68\%) \\
          &    & transport, storage and communication (9.63\%) \\
          &    & wholesale/retail (11.23\%) \\
          &    & other community, social and personal service (9.63\%) \\
\\
legal status & LS & sole proprietorship (4.88\%) \\
             &    & partnership (10.57\%) \\
             &    & limited liability partnership (7.50\%) \\
             &    & limited liability company (77.05\%) \\
\\
startups & SU & no (97.5\%) \\
         &    & yes (2.5\%) \\
\\
London \& South East & LS & no (23.61\%) \\ 
                    &    & yes (76.39\%) \\
\bottomrule
\end{tabular}
\end{table}

We chose the SMEs dataset for two main reasons. First, SMEs play a crucial role in national economic development, making it vital for banks and financial institutions to provide essential support and ensure fair treatment. An algorithmic fairness auditing tool is, therefore, essential for financial regulators. It also serves as a self-assessment resource for lenders, helping them ensure compliance with the recent EU AI Act requirements in their AI decision systems during product development. However, due to limited access to SMEs' data, there is a significant research gap regarding algorithmic bias in SMEs' access to finance. Second, the SMEs dataset is survey-based and characterised by relatively low quality, making it an ideal choice for stress-testing our proposed framework to assess its effectiveness in handling such data challenges.

To avoid redundancy, we selected survey results from 2012Q4 to 2020Q2 and focused on 15 important features identified in the literature \citep{sun_what_2021, calabrese_expectations_2022, cowling_access_2016, cowling_has_2022, cowling_small_2012} (see Table~\ref{tab:description_dataset} for details). These features capture various aspects of the loan application process. After filtering out data points with more than 20\% missing features, the final dataset comprised 4,159 entries for analysis. Details of the data cleaning process can be found in Supplementary Materials.

To apply our proposed framework, we consider the 15 features listed in Table~\ref{tab:description_dataset} as \(X\), and treat the firm size as the protected attribute \(S\), defined by a combination of the number of employees and annual turnover (Micro-firms are defined as those with fewer than 10 employees and an annual turnover of less than £2 million following the literature \citep{sun_what_2021}). Such grouping leads to 1,719 micro-firms (\(s = s_{-}\))  and 2,440 non-micro firms (\(s = s_{+}\)) as non-protected group. Our dataset does not exhibit a significant imbalance in the protected attributes, which is useful, as this would complicate testing our framework's universality in different imbalance levels as in Section~\ref{sec:imbalance}. Oversampling to adjust imbalance could alter feature relationships, making bias auditing unreliable \citep{chen_interpretable_2024}. Instead, we maintain a moderate imbalance and vary the imbalance level by under-sampling.
For the target variable, we use the outcome of bank loan application,  due to the significant role of bank loans in SME financing \citep{sun_what_2021}. The dataset records 3,391 approvals (\(y = 1\)) and 768 rejections (\(y = 0\)), highlighting the decisions faced by SMEs in access to finance.

Following the workflow outlined in \autoref{fig:framework}, we focus on the 1,719 micro-firms to determine whether they have experienced algorithmic bias.
As described in the workflow, we use logistic regression as the default model for both prediction and fitting for simplicity (see performance evaluation and robustness tests in Supplementary Materials). 
For the fitting and prediction, the data are typically split into training (80\%) and testing (20\%) sets, with hyper-parameters optimised via grid search and 5-fold cross-validation. The model yielding the highest AUC value is selected for predictions on the target \(Y\).
Without loss of generality, we set the default \(\delta\) to 0.3 times the standard deviation of the micro-firms' \emph{IC}s. This flexible threshold can be adjusted according to the specific dataset and research context. Additional robustness tests regarding threshold adjustments are provided in Supplementary Materials.
The remaining settings are \(N = 100\) and \(K = 30\). We include only micro-firms with more than 35 peers to meet the basic requirement for a large sample size. Due to the limitations of our dataset quality, firms with fewer than 35 peers are labelled as ``Unknown'' and, for illustrative purposes, will not be included in our further analysis. However, for real auditing tasks, the dataset size and quality are typically higher than those obtained from surveys, and this issue is likely to be mitigated.
For the hypothesis testing step, the default test is \(H_0\) vs. \(H_1\) with a significance level of 5\%. However, to differentiate between cases of discrimination and privilege, we also run tests for \(H_2\) and \(H_3\) against their respective alternative hypotheses. These tests compare the mean approval likelihood of the peers against that of the micro-firms, thereby identifying potential algorithmic bias in terms of discrimination or privilege.

\section{Results}
\label{sec:results}
In this section, we present the experimental results on the SMEs dataset, demonstrating the efficacy of our ``peer-induced fairness'' framework.

\subsection{Algorithmic fairness auditing}
\label{sec:algorthmic_discovery}

\begin{figure}[H]
        \centering
        \includegraphics[width=0.45\textwidth, height=0.25\textheight]{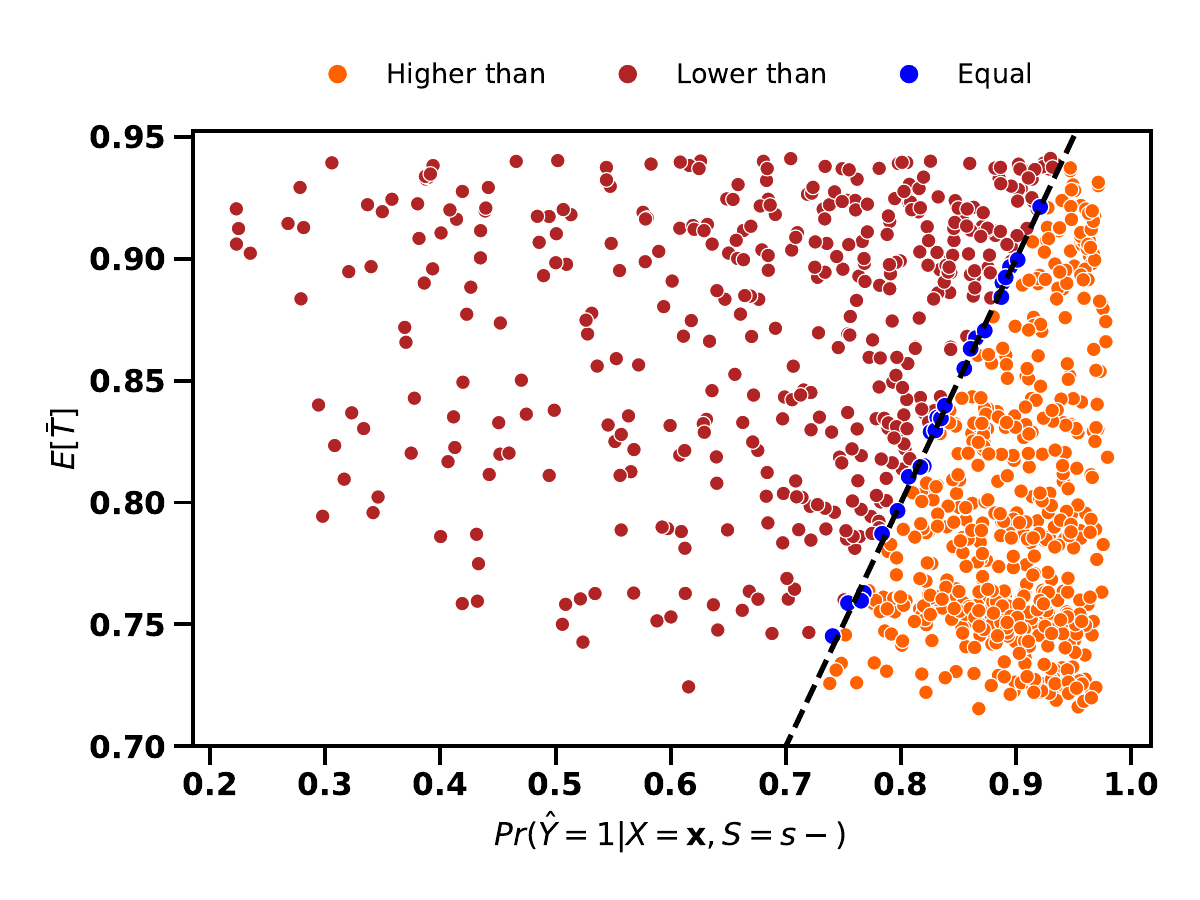}
  \caption{Comparative analysis of loan approval likelihood for micro-firms against peers. The black dashed 45-degree line, denoting $Y=X$, symbolises perfect fairness. Red and orange data points represent micro-firms with approval likelihoods significantly lower or higher, respectively than the average of their peers. Blue points denote no significant difference.}
  \label{fig:visualisation_and_assessment}
\end{figure}

Following the workflow outlined in \autoref{fig:framework} and the experimental settings described in Section~\ref{sec:experiment}, we successfully identified algorithmic bias within the SMEs dataset. The scatter plot in Figure~\ref{fig:visualisation_and_assessment}, which compares approval likelihoods between micro-firms and their peers, reveals that only 2.48\% of micro-firms are treated fairly, indicating significant disparities in the credit approval system. The remaining 97.52\% experience algorithmic bias, with 41.51\% of micro-firms facing discrimination. Interestingly, 56.40\% of micro-firms, despite being underrepresented, benefit from the decision system by receiving approval likelihoods higher than the average of their peers.

\begin{figure}[H]
        \centering
        \includegraphics[width=\textwidth, height=0.5\textheight]{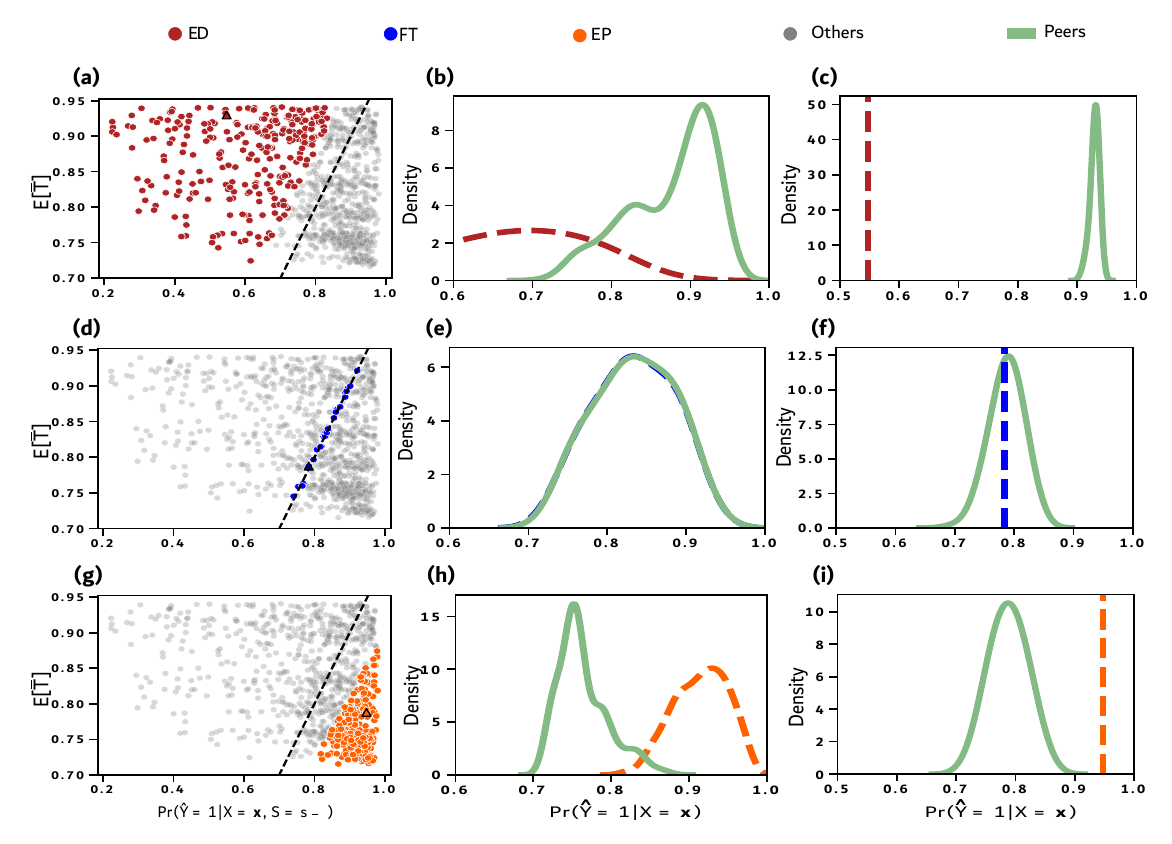}
\caption{Comparative analysis of loan approval likelihood for micro-firms under each algorithmic treatment category against peers. (a)-(c): Extremely discriminated (ED) micro-firms. (d)-(f): Fairly treated (FT) micro-firms. (g)-(i): Extremely privileged (EP) micro-firms. The coloured data points in the first column of each row represent a comparison among peers within each category. The x-axis shows the approval likelihood for micro-firms, while the y-axis displays the average approval likelihood of the peers. The second column compares the approval likelihood between these micro-firms (i.e., coloured points in (a), (d), (g)) and their peers at the group level. The third column provides the comparison, at the individual level, between the selected micro-firm (i.e., coloured triangle in (a), (d), (g)) and its peers.}
  \label{fig:category_visualization}
\end{figure}

To identify the specific extent of discrimination and privilege faced by each micro-firm, we compare the approval likelihood difference between a given micro-firm $A=(s-,\bm{x_0})$ and its peers. For micro-firms with a higher likelihood of approval, we allow for greater tolerance when assessing extreme algorithmic bias, adjusting the standard based on each firm's approval likelihood. Specifically, we consider a micro-firm to experience extreme algorithmic bias if the absolute difference exceeds 0.1 times its own approval likelihood. Mathematically, this is expressed as $|\mathbb{P}(\hat{Y}_{s_{-}}| s_{-}, \bm{x}_0)-\mathbb{E}[\bar{T}_i]| > 0.1 \times \mathbb{P}(\hat{Y}_{s_{-}}| s_{-}, \bm{x}_0)$. A negative difference indicates discrimination, while a positive difference signifies privilege. This approach ensures the flexibility of the standard, making it suitable for firms in different situations. Instead, if the absolute difference is less than the threshold, it represents slight discrimination or slight privilege. Further details can be found in the Supplementary Materials. In certain cases, such slightly unfair treatment may be considered fair, depending on the regulatory tolerance and the specific industry being audited. 

Specifically, in our case, 26.71\% of micro-firms experience extreme discrimination, with their approval likelihood markedly lower than that of their peers, as shown in Figure~\ref{fig:category_visualization} (a)-(c), at both group and individual levels. 32.17\% of micro-firms are extremely privileged, as shown in Figure~\ref{fig:category_visualization} (g)-(i). Even though algorithmic privilege might seem beneficial for micro-firms, neither scenario is desirable. We advocate for transparency and fairness in decision-making processes. Arbitrary or opaque factors influencing decisions are contrary to the principles of fairness and should be rigorously addressed to ensure equitable treatment across all applicants.

It is important to emphasise that our framework is a tool for audits by regulators and stakeholders, aiming to detect algorithmic bias. In credit loan applications, rejected customers are particularly concerned about whether they were rejected and discriminated against, while regulators and banks require detailed results to audit the fairness of their models for all applicants. Therefore, our framework also includes detailed information on accepted applicants. Additionally, without compromising generalisation to other research areas, it is crucial to focus on all applicants.

\begin{figure}[H]
    \centering
        \includegraphics[width=0.5\textwidth, height=0.28\textheight]{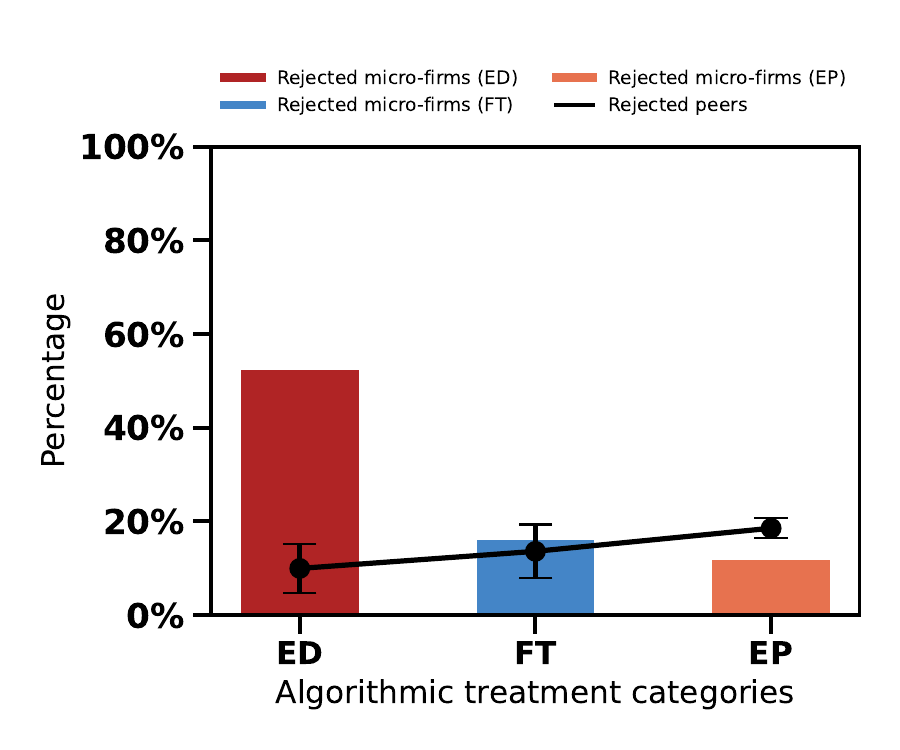}
  \caption{Rejection rates of micro-firms across algorithmic treatment categories and their peers. The algorithmic treatment categories include extremely discriminated (ED), fairly treated (FT), and extremely privileged (EP). Each category includes multiple micro-firms with a single rejection rate, shown as histograms, while the rejection rate of peers of each micro-firm in this category is represented in the black line with error bars to indicate variability.}   \label{fig:discrimination_and_outcomes}
\end{figure}

We also validate our framework by investigating the connection between accessing finance outcomes and disparities in algorithmic bias. Among these markedly discriminated micro-firms, 52.42\% were denied loans, whereas only 9.97\% of their peers faced rejection, highlighting a significant disparity in rejection rates. The rejection rate of micro firms decreases and that of their peers increases with the diminished discrimination. The difference in rejection rates between micro-firms and their peers also decreases. The rejection rates of peers fluctuate around the rejection rate of fairly treated micro-firms. This fluctuation indicates that within the category, some micro-firms experience higher rejection rates compared to their peers, while others experience lower rejection rates, illustrating a gradual convergence in rejection rates across categories with less pronounced discrimination. Notably, even the lowest peer rejection rate at the bottom of the error bar surpasses that of micro-firms in the extremely privileged category, where micro-firms experience the lowest rejection rates, as in Figure~\ref{fig:discrimination_and_outcomes}. These findings, derived from our bias audit based on financing outcomes prediction, align with the observed financing results. This congruence further validates the utility of our framework in accurately reflecting disparities and biases in the loan approval process. Further details on the extent of algorithmic bias are provided in Supplementary Materials, which expands the analysis to include two additional categories: slightly discriminated and slightly privileged. The analysis shows that even with these detailed treatment categories, the results consistently validate the effectiveness of our framework.

The above experimental results reveal that our ``peer-induced fairness'' framework not only effectively identifies disparities in algorithmic fairness but also facilitates the visual representation of individual-level discrepancies across all users in the dataset. This capability enables clear visualisation of algorithmic fairness, making discrimination or privilege readily distinguishable. Such insights are invaluable for both regulatory purposes and for verifying the effectiveness of algorithmic fairness models.

\subsection{Data scarcity and imbalance}
\label{sec:imbalance}

Data scarcity and imbalance significantly influence the performance of advanced machine learning models due to the potential for inaccurate parameter estimation \citep{wang2024representative}. This issue is especially pronounced in many datasets, where the representation of minority groups is often limited compared to majority groups \citep{chen_interpretable_2024,lessmann_benchmarking_2015}. This discrepancy caused by the poor data quality, subsequently affects the auditing of algorithmic bias.

Our ``peer-induced fairness'' framework addresses these challenges uniquely. Unlike traditional models that rely heavily on the data from the protected group, our framework bases all parameter estimations on peers identified within the unprotected group. This group typically possesses ample data points, effectively mitigating issues related to data scarcity and group imbalance, making our framework robust theoretically. 

We investigate the stability and credibility of our ``peer-induced fairness'' framework by evaluating the percentages of unfairly treated ($PUT$) protected individuals or organisations and the invariant outcome ratio ($IOR$) under varying levels of imbalance.
The imbalance ratio, \(\omega\), is defined as the proportion of samples in the protected class:
\begin{equation}
    \omega = \frac{ \# (S = s_{-}) }{ \# (S = s_{+}) + \# (S = s_{-}) },
\end{equation}
where \(\# (\cdot)\) denotes the cardinality of a set. A perfectly balanced dataset corresponds to \(\omega = 50\%\).
The $PUT$ is calculated as the number of unfairly treated individuals or organisations divided by the total number of selected subjects in the experiments with different \(\omega\). 
The $IOR$ is computed as the number of selected individuals or organisations in the experiment with \(\omega\) that have unchanged predictive outcomes compared to the original experiment divided by the number of commonly selected subjects in both the experiment with \(\omega\) and the original experiment.

In the SMEs experiment, building upon the default settings outlined in Section~\ref{sec:experiment}, we investigate the impact of varying imbalance ratios by randomly selecting subsets of the original dataset with controlled imbalance levels. Specifically, we evaluate the framework's performance at imbalance ratios of \(\omega = \{36.33\%, 31.33\%, 26.33\%, 21.33\%, 16.33\%, 11.33\%\}\), where the original dataset's imbalance ratio is \(\omega = 41.33\%\). By gradually decreasing the proportion of micro-firms in these subsets, we assess the framework's robustness across different levels of imbalance. To minimise the effects of randomness in subset selection, this process is repeated five times. The detailed procedure is provided in Supplementary Materials.

\begin{figure}[H]
        \centering
        \includegraphics[width=0.5\textwidth, height=0.25\textheight]{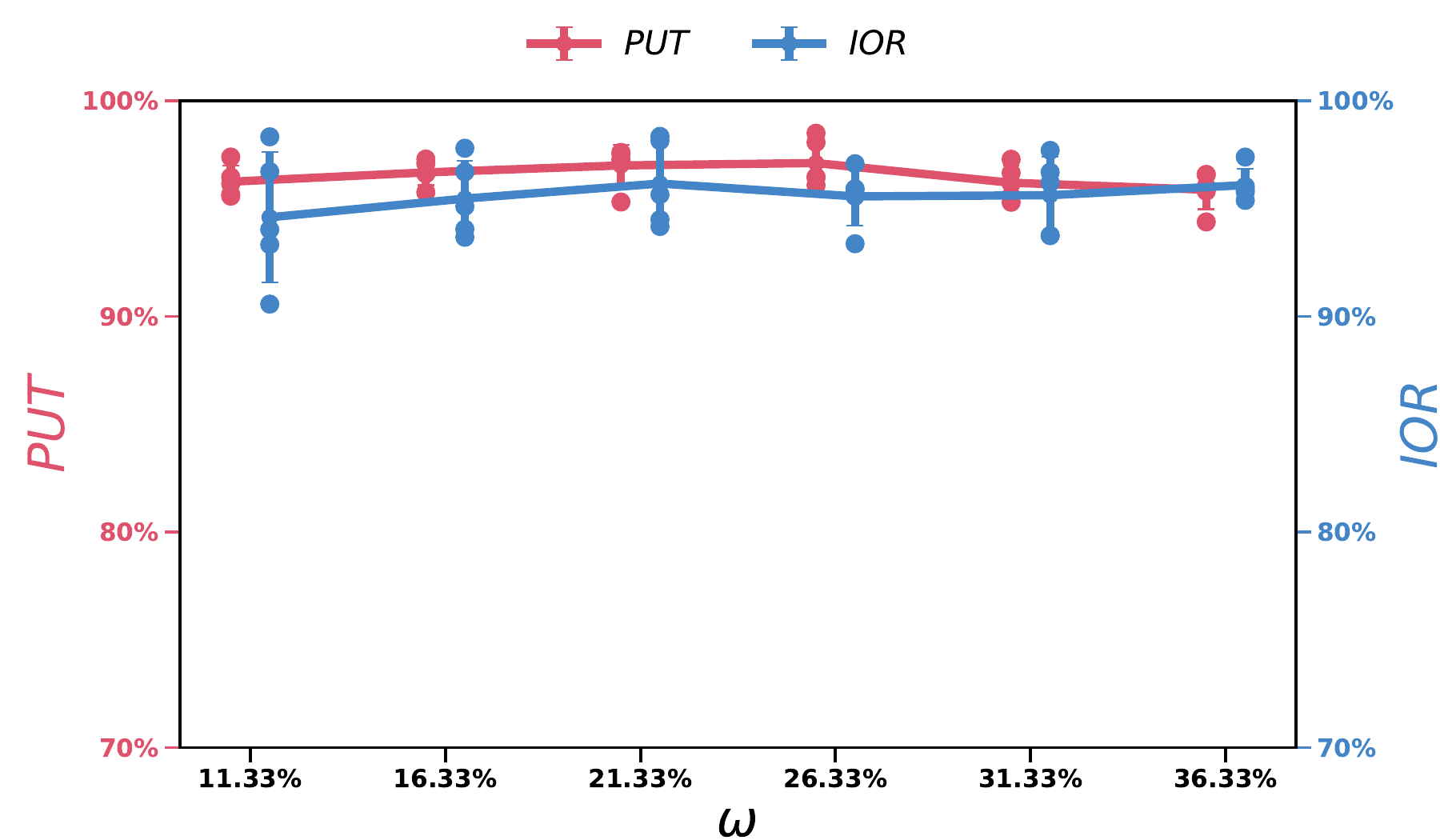}
        \caption{Percentage of unfairly treated micro-firms and invariant outcome ratio at different group imbalance levels. The imbalance level is represented on the x-axis as a percentage, ranging from 11.33\% to 36.33\%. The left y-axis shows the percentage of unfairly treated micro-firms (blue line), while the right y-axis displays the invariant outcome ratio (red line) as the imbalance level changes from the initial level to other levels.}
        \label{fig:imbalance_proportion}
\end{figure}

The results are visualised in Figure~\ref{fig:imbalance_proportion} and demonstrate the robustness of our framework.
From the view of $PUT$, the small error bars across all the imbalance levels suggest the results across the five repetitions are highly consistent. This observation underscores the robustness of our ``peer-induced framework'' to imbalanced datasets. From the view of $IOR$, it is approximately 95\% and remains stable across different imbalance levels. This aligns with our expectations, as the framework does not rely on data from the minority group but rather leverages information from the unprotected group, leading to inherent robustness. The small error bars also suggest that the results regarding $IOR$ in these five repeats are highly consistent.

These findings highlight the universality of our ``peer-induced fairness'' framework with respect to the different data quality, as the auditing results remain consistent despite variations in imbalance levels. This distinguishes our framework from others by effectively addressing the prevalent challenges of data scarcity and imbalance in the field. Regulators can utilise this framework to evaluate the practices of companies and institutions, while these organisations can also reliably employ it for thorough self-assessment. Additionally, an alternative computation method is detailed in Supplementary Materials to further enhance the robustness of our approach.

\subsection{Explainable fairness discovery}
\label{sec:explanation}
Next, we focus on the final step outlined in \autoref{fig:framework}, which involves providing explanations for individuals who are fairly treated but receive an outcome of \(y = 0\). Our explanation approach is based on comparing the features of these individuals with those of their peers. This method helps avoid misunderstandings and ambiguity, allowing us to offer a clear ``watch-out'' list of features that may have contributed to the unfavourable decision.

\begin{figure}[H]
    \centering
    \includegraphics[width=0.8\textwidth, height=0.24\textheight]{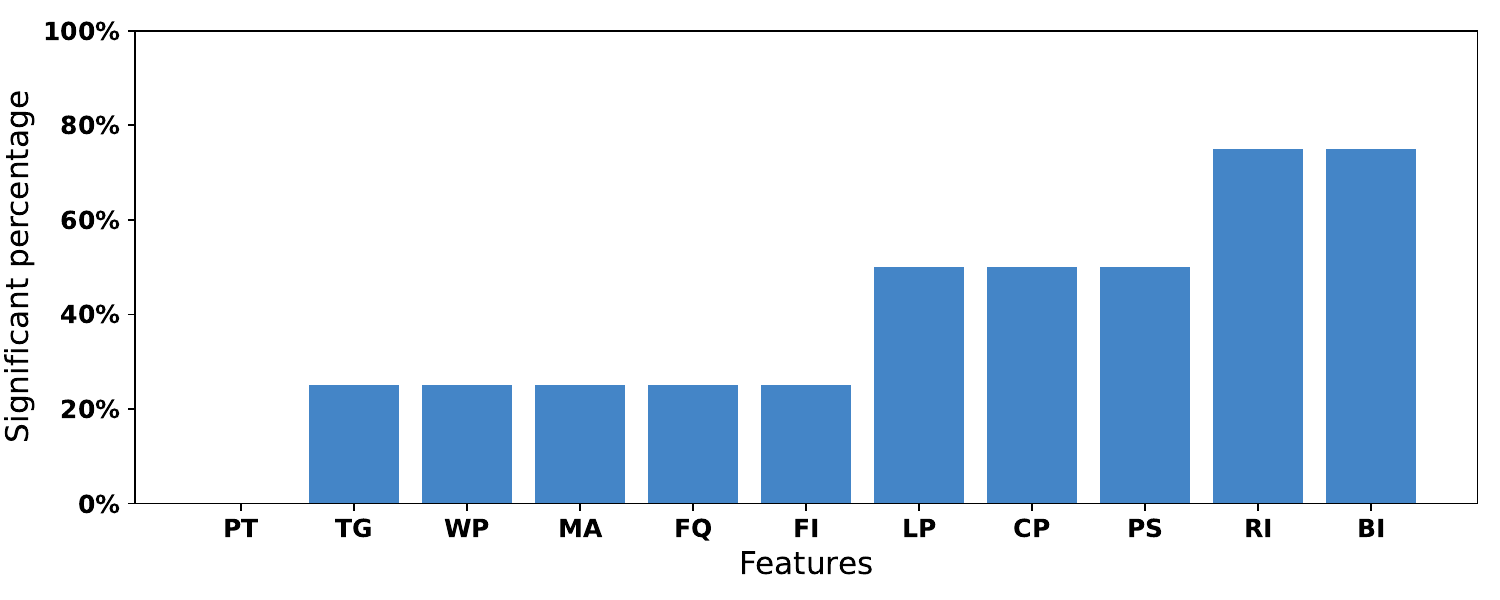}
    \caption{Comparative analysis of non-intrinsic features for rejected while fairly treated micro-firms vs. accepted peers. The x-axis represents the selected key attributes being analysed, including finance qualification for manager (FQ), written plan (WP), previous turn-down (PT), loss or profit (LP), risk (RI), product/service development (PS), business innovation (BI), regular management account (MA), turnover growth rate (TG), credit purchase (CP) and funds injection (FI). The y-axis represents the percentage of those micro-firms with significantly worse performance than their accepted peers on each feature.}
    \label{fig:significant_percentage}
\end{figure}

Given the existence of accepted peers as the counterfactual instances with positive accessing finance outcomes, the micro-firm which is fairly treated should originally have the same outcomes. 
Our framework identifies the feature differences between each rejected while fairly treated micro-firm and its accepted peers by another hypothesis testing, with details presented in Supplementary Materials.
For each feature, we summarise the percentage of these micro-firms that perform significantly worse than their accepted peers. We consider some non-intrinsic features to identify and understand these discrepancies, as in Figure~\ref{fig:significant_percentage}. The descriptions for each feature value are shown in Supplementary Materials. Results show that even though none of them have been rejected previously and only 25\% of them perform worse on financial qualifications and written plans, banks generally prioritise the financial and business health of firms. 75\% of these micro-firms invest excessively in business innovation and have lower risk ratings. Besides, half of them invest in product/service development and have lower profits. The uncertain returns and high risks associated with innovation lead to the failure or commercial non-viability of most innovative products \citep{coad2008innovation, Hall2002financing, freel2007small}, exacerbating already poor-performing risk indicators. The worse performance on these key features makes banks cautious about the long-term financial sustainability of these firms. It also reflects the capability of these micro-firms, negatively affecting their loan approvals.

This exploration identifies the differences between micro-firms and their peers for each feature and summarises the percentage of micro-firms that perform worse on each feature. This explainable analysis not only enhances the transparency of our framework but also supports regulators and stakeholders in understanding the specific challenges most incapable micro-firms face, and highlights the features that they need to watch out for and pay extra attention to.

\section{Concluding remarks and discussion}\label{sec:conclusion}

In the age of AI, where automated decision-making systems increasingly determine access to essential services such as finance, housing, and employment, the consequences of algorithmic bias can be severe. Discriminatory practices not only undermine social equity but also violate legal and ethical standards, potentially causing significant harm to vulnerable groups.  Various regulatory documents have emphasised the need for ongoing oversight and auditing of decision systems \citep{voigt_eu_2017, madiega2021artificial, BSI2023}, both at the initial deployment stage and throughout their operational life cycle \citep{madiega2021artificial}. This focus on fairness is not confined to the European Union; several countries, including the United Kingdom and the United States \citep{president2016big, BSI2023}, have also introduced regulations and guidelines to ensure that AI and automated decision-making systems function transparently and equitably. 

To meet the growing regulatory demands across, we proactively and timely introduce an algorithmic fairness auditing framework. It is a robust auditing framework for both internal and external assessment in a plug-and-play fashion. Designed as a fully modular tool, the framework allows users—whether financial institutions, regulators, or third-party auditors—to customise settings based on their specific needs and objectives, making it highly adaptable across various sectors. The core idea is grounded in peer comparisons, which is both intuitive and intrinsic. This approach is computationally efficient and robust to varying data qualities, ensuring reliable auditing results in different scenarios. By facilitating comprehensive fairness audits, our tool helps prevent automated systems from perpetuating or exacerbating existing inequalities. Our framework also enhances transparency by providing necessary explanations and ``watch-out'' lists for those who receive unfavourable decisions due to insufficient capabilities. This feature promotes understanding and trust among users and affected groups, aligning with regulatory requirements for transparency and accountability. In a regulatory landscape where continuous monitoring and auditing are increasingly mandated, such tools are indispensable. They offer a practical means to ensure that AI systems are both legally compliant and socially responsible, adhering to broader ethical imperatives for fairness and accountability. 

From an empirical standpoint, our framework uncovers alarming issues in the current state of SMEs' access to finance. Specifically, our findings indicate that only 2.48\% of micro-firms are treated fairly, while a staggering 41.51\% face discrimination. Even when we adjusted the data quality by altering the imbalance level, the audit results remained highly consistent with our original findings. These results underscore a serious and inequitable banking environment that demands immediate attention. Additionally, we observed that some micro-firms are rejected due to inherent limitations, such as higher risk or greater investment in innovation, rather than discrimination when compared to their peers. These empirical findings also demonstrate the effectiveness and robustness of our framework in real-world applications. For these reasons, we believe our framework represents a significant advancement and a policy-relevant contribution to algorithmic fairness.

Given the modular structure of our framework, there is significant potential for further enhancement and adaptation. Currently, our framework is based on a static causal model, which, while effective for many applications, may not fully capture the complexities of real-world scenarios where dynamic causal models are more appropriate. In such cases, feedback loops can alter relationships over time, as decisions made based on certain features can influence future data and outcomes. A static framework may not adequately account for these evolving interactions. However, the core concept of peer comparison remains valid even within a dynamic causal model. Future studies could focus on integrating dynamic causal modelling into our framework to better address these feedback mechanisms, ensuring its applicability and robustness across a broader range of contexts.

\section*{Data availability}
Data and codes are available at \href{https://doi.org/10.5255/UKDA-SN-6888-26}{UK Data Archive} and \href{https://github.com/99Catherine99/peer-induced-fairness/}{GitHub} respectively.

\section*{Acknowledgements}
The authors of this manuscript would like to thank Prof.Raffaella Calabrese and Dr.Yizhe Dong for their assistance and support in the discussion and research direction.

\begin{appendices}
\renewcommand{\thesection}{Appendix \Alph{section}}

\section{Proof of Theorem~\ref{thm:delta-peer-identification}}
\label{sec:appendix_proof_thm}
\begin{proof}
    According to Definition~\ref{def:delta-peer}, we have
    \begin{align}
        &\quad \left| \mathbb{P}(\mathcal{G}(s_{+}, \bm{x}_j)) - \mathbb{P}(\mathcal{\tilde{G}}(s_{+}, \bm{x}_0)) \right|  \nonumber \\
        &= \Big| \mathbb{P}(S, \bm{X}(s_{+}), Y(s_{+}, \bm{x}_j)) - \mathbb{P}(S, \bm{X}(s_{-}), Y(s_{+}, \bm{x}_0)) \Big| \nonumber \\ 
        &= \mathbb{P}(s_{+}) \cdot \Big|  \mathbb{P}(\bm{X}(s_{+})) \cdot \mathbb{P}(Y(s_{+}, \bm{x}_j))  - \mathbb{P}(\bm{X}(s_{-})) \mathbb{P}(Y(s_{+}, \bm{x}_0))   \Big|  \nonumber \\
        &= \mathbb{P}(s_{+}) \cdot \Big| \mathbb{P}(\bm{x}_j) \cdot \xi(s_{+}, \bm{x}_j) \cdot \mathbb{P}(Y |s_{+}, \bm{x}_j)  - \mathbb{P}(\bm{x}_0) \cdot \xi(s_{-}, \bm{x}_0) \cdot \mathbb{P}(Y |s_{+}, \bm{x}_0) \Big| \nonumber \\
        &= \mathbb{P}(s_{+}) \cdot \Big| \xi(s_{+}, \bm{x}_j) \cdot \mathbb{P}(Y, \bm{x}_j|s_{+})  - \xi(s_{-}, \bm{x}_0) \cdot \mathbb{P}(Y, \bm{x}_0 |s_{+}) \Big| \nonumber \\
        &= \mathbb{P}(s_{+}) \cdot \mathbb{P}(Y, \bm{x}_j | s_{+}) \cdot|\xi(s_{-}, \bm{x}_0) - \xi(s_{+}, \bm{x}_j) | \nonumber \\
        &\leq \mathbb{P}(s_{+}) \cdot \mathbb{P}(Y, \bm{x}_j | s_{+}) \cdot \delta \nonumber \\
        &< \delta. \nonumber
    \end{align}
The derivation of the second equation is underpinned by the factorisation property, as detailed in Eq.~\eqref{eq:G(s, x)} and Eq.~\eqref{eq:CG(s+, x)}. 
The transition to the third equation leverages the modularity property, which is articulated in Eq.~\eqref{eq:Y(s,x)}. The transition from $\mathbb{P}(\bm{X}(s_{+}))$ and $\mathbb{P}(\bm{X}(s_{-}))$ into $\mathbb{P}(\bm{x}_j) \cdot \xi(s_{+}, \bm{x})$ and $\mathbb{P}(\bm{x}_0) \cdot \xi(s_{+}, \bm{x}_0)$ refer to Eq.~\eqref{eq:P(X) = x}.
Regarding the fifth equation, it addresses the practical consideration of dealing with high-dimensional continuous variables in \(\bm{X}\). 
Given the high-dimensional nature of \(\bm{X}\), the probability of \(\bm{X}\) equating to a specific value within this space is nominally small. 
Thus, for practical purposes, the distinction between \(\mathbb{P}(Y, \bm{X} = \bm{x}_j|s_{+})\) and \(\mathbb{P}(Y, \bm{X} = \bm{x}_0 |s_{+})\) is considered negligible (i.e., $\mathbb{P}(Y, \bm{x}_j | s_{+})=\mathbb{P}(Y, \bm{x}_0 |s_{+})$). 
Therefore, $C$ is considered as a peer of $A$ according to Definition~\ref{def:delta-peer}.
\end{proof}

\section{Implementation for peer identification}
\label{sec:appendix_algo}

\begin{algorithm}[H]
\caption{Identification of $\delta$-Peers for Protected Individuals}\label{algo}
\begin{algorithmic}[1] 
    \Require A set of individuals $\{A\} = \{(s_{-}, \bm{x}_0)\}$ from the protected group, a set of individuals $\{B_i\}_{i=1}^{N} = \{(s_{+}, \bm{x}_i)\}$ from the unprotected group, a threshold $\delta$, and a minimum number of peers $U$.
    \Ensure A subset of $\{B_i\}$ designated as $\delta$-peers of $A$, with each protected individual having at least $U$ peers.

    \ForAll{$A = (s_{-}, \bm{x}_0)$}
        \State Initialise an empty list of peers for $A$, denoted as $\text{Peers}_A$
        \State Compute $\xi(s_{-}, \bm{x}_0)$ for $A$
        \ForAll{$B_i = (s_{+}, \bm{x}_i)$ in $\{B_i\}_{i=1}^{N}$}
            \State Compute $\xi(s_{+}, \bm{x}_i)$ for $B_i$
            \State Calculate the difference $\Delta = |\xi(s_{-}, \bm{x}_0) - \xi(s_{+}, \bm{x}_i)|$
            \If{$\Delta < \delta$}
                \State Add $B_i$ to $\text{Peers}_A$
            \EndIf
        \EndFor
    \EndFor
\end{algorithmic}
\end{algorithm}

\section{Proof of Proposition~\ref{prop:synthetic}}
\label{sec:appendix_proof_prop}

\begin{proof}
    To demonstrate that the synthetic individual \(\bar{T}_i\) qualifies as a \(\delta\)-peer of \(A\), we compare \(A\)'s \emph{IC}, \(\xi(s_-, \bm{x}_0)\), against the average \emph{IC} of any \(K\) peers of \(A\), denoted as \(\sum_{j=1}^{K} \xi(s_+, \bm{x}_j) / K\). The difference is calculated as follows:

\begin{align*}
    & \quad \left|\xi(s_-, \bm{x}_0) - \frac{1}{K} \sum_{j=1}^{K} \xi(s_+, \bm{x}_j)\right| \\
    & = \frac{1}{K} \left|K\xi(s_-, \bm{x}_0) - \sum_{j=1}^{K} \xi(s_+, \bm{x}_j) \right| \\
    & = \frac{1}{K} \left|(\xi(s_-, \bm{x}_0) - \xi(s_+, \bm{x}_1))  + \cdots + (\xi(s_-, \bm{x}_0) - \xi(s_+, \bm{x}_K)) \right| \\
    & \leq \frac{1}{K} \sum_{j=1}^K \left|\xi(s_-, \bm{x}_0) - \xi(s_+, \bm{x}_j)\right| \\
    & \leq \delta.
\end{align*}

This inequality shows that the average discrepancy between \(A\)'s \emph{IC} and that of \(\bar{T}_i\) is within \(\delta\). Hence, according to Theorem~\ref{thm:delta-peer-identification}, \(\bar{T}_i\) indeed qualifies as a \(\delta\)-peer of \(A\).
\end{proof}

\end{appendices}

\bibliographystyle{apalike}
\bibliography{reference} 

\end{document}